\documentclass[preprint,aps,superscriptaddress,floats,showpacs]{revtex4}
\usepackage{graphicx}
\usepackage{amsmath}
\usepackage{amsfonts}
\usepackage{amssymb}
\usepackage{epsfig}
\usepackage{color}
\usepackage{bm}

\begin{document}

\title{Generation of high-charge electron beam in a subcritical-density plasma through laser pulse self-trapping}

\author{V.~Yu.~Bychenkov}
\affiliation{P. N. Lebedev Physics Institute,
	Russian Academy of Science, Leninskii Prospect 53, Moscow 119991,
	Russia}
\affiliation{Center for Fundamental and Applied Research,
Dukhov Research Institute, Moscow 127055, Russia}
\author{M.~G.~Lobok}
\affiliation{P. N. Lebedev Physics Institute, Russian Academy of
	Science, Leninskii Prospect 53, Moscow 119991, Russia}
\affiliation{Center for Fundamental and Applied Research,
Dukhov Research Institute, Moscow 127055, Russia}
\author{V.~F.~Kovalev}
\affiliation{Keldysh Institute of Applied Mathematics, Russian Academy of Sciences, Moscow 125047, Russia}
\affiliation{Center for Fundamental and Applied Research,
Dukhov Research Institute, Moscow 127055, Russia}
\author{A.~V.~Brantov}
\affiliation{P. N. Lebedev Physics Institute,
Russian Academy of Science, Leninskii Prospect 53, Moscow 119991,
Russia}
\affiliation{Center for Fundamental and Applied Research,
	Dukhov Research Institute, Moscow 127055, Russia}

\begin{abstract}
To maximize the charge of a high-energy electron beam accelerated by an
ultra-intense laser pulse propagating in a subcritical plasma, the pulse
length should be longer than both the plasma wavelength and the laser
pulse width, which is quite different from the standard bubble regime. In
addition, the laser--plasma parameters should be chosen to produce the
self-trapping regime of relativistic channeling, where the diffraction
divergence is balanced by the relativistic nonlinearity such that the laser
beam radius is unchanged during pulse propagation in a plasma over many
Rayleigh lengths. The condition for such a self-trapping regime is the same
as what was empirically found in several previous simulation studies in the
form of the pulse width matching condition. Here, we prove these findings
for a subcritical plasma, where the total charge of high-energy electrons
reaches the multi-nC level, by optimization in a 3D PIC simulation study and
compare the results with an analytic theory of relativistic self-focusing. A
very efficient explicitly demonstrated generation of high-charge electron
beams opens a way to a high-yield production of gammas, positrons, and
photonuclear particles.
\end{abstract}

%\pacs{ 52.38.Kd, 52.65.Rr, 29.25.Ni}

\maketitle

\section{Introduction}\label{sec1}

A laser-excited plasma wakefield \cite{Tajima} is a promising way to
accelerate electrons to high energies for a source in gamma and
photonuclear applications. The best-known 3D wakefield structure, the
so-called bubble, which has been observed in PIC (particle-in-cell)
simulations and in experiments, is stable in a rarefied plasma with an
electron density $n_{\mathrm{e}}\lll n_{\mathrm{c}}$ ($n_{\mathrm{c}}$ is
the critical density) for laser pulses shorter than or of the order of the
plasma wavelength, $L\lesssim\lambda_{\mathrm{p}}$, and shorter than the
laser pulse width, $L<d$ \cite{Pukhov}. The main aim of those
investigations was to develop a source of high-quality particle beams with
energies of many hundred MeV or GeV, a high spatial quality, and a
monoenergetic energy distribution for practical applications. But the total
charge of high-energy electron beams is typically at the multi-pC level. On
the other hand, there are practical applications that do not require such
high electron energies and beam quality but do require a much higher
quantity of accelerated electrons with an average energy of the order of
$\sim100$\,MeV. This work reports and justifies how this can be done.

We recently performed a preliminary study to determine the optimal density
and thickness of planar low-density targets maximizing the number of
high-energy electrons generated by a femtosecond laser pulse with a given
intensity \cite{Lobok}. The interest in such a study is related to laser
generation of electron bunches that can produce hard $\gamma$-quanta
suitable for radiography of dense thick samples with $\gamma$-energy of a
few MeV, electron--positron pairs, different ($\gamma,n$) reactions for
neutron generation, and even light mesons. To be of practical significance,
these applications require a high total charge of accelerated electron
bunches considerably exceeding standard rarefied gas densities (used for
wakefield/bubble acceleration) \cite{Yang,Willingale,Goers}. The considered
acceleration regime is quite different from the standard bubble regime with
conditions opposite to those in Ref.~\cite{Pukhov}, $L>\lambda_{\mathrm{p}},
\,d$, \cite{Lobok} and occurs when the nonlinear 3D charge-separation
structure travels as a stable elongated laser field-filled cavity over many
Rayleigh lengths in a dense gas plasma, more like a ``laser bullet'' than an
empty bubble.

Here, we advance the study of a ``light bullet'' propagating in a
near-critical plasma. An advantage of such a dense plasma is that it allows
a relativistically strong laser pulse with a modest pulse energy to produce
an electron bunch with a high charge (multi-nC) in the hundred-MeV energy
range because the pulse has a short duration and is tightly focused. The
peculiarity of electron acceleration is associated with a cumulative effect
of direct laser acceleration (DLA) and wakefield acceleration \cite{Shaw}
and bears the features of stochastic acceleration \cite{Lobok}. Using a 3D
PIC model (Sec.~\ref{sec2}), we investigate the interplay of the laser pulse
and plasma parameters in detail to best choose them in terms of maximizing
the total charge of the generated electron bunches with energies in the
hundred-MeV range. This maximization requires stable propagation of the
laser pulse for a long distance, which occurs under the empirically found
condition (in PIC simulations) of matching the self-consistent waveguide
radius (laser cavity radius) to the electron plasma density and laser
intensity \cite{gordienko,lu,ralph}. In Sec.~\ref{sec3}, we present a
detailed study of laser pulse propagation and electron generation under the
matching condition for laser cavity radii with different laser parameters.
This matching condition is shown to be the condition for a self-trapping
regime of relativistic self-channeling, which we prove in Sec.~\ref{sec4}
using a recently developed theory \cite{kovalev}. We discuss the results and
conclusions in~Sec.~\ref{sec5}.

\section{Simulation model}\label{sec2}

We studied laser pulse propagation in an underdense planar target and
generation of electron bunches by a laser-generated space-charge structure
using 3D PIC simulations with the high-performance electromagnetic code VSim
(VORPAL). We considered the interaction of a linearly polarized laser pulse
(in the $z$ direction) with the wavelength $\lambda=2\pi c/\omega=1\,\mu$m
of variable energy and a Gaussian intensity shape ($I$) both in time (with
an FWHM duration $\tau=30$\,fs) and in space, $I=I_0\exp[-x^2/R_L^2]$, where
$R_L=2.4\,\mu$m. The laser pulse propagated in a transparent subcritical
plasma target consisting of electrons and ions with an atomic mass to charge
ratio equal to two. For sake of the maximum accuracy we are not limited to the case of immobile ions.
The normalized laser field amplitude $a_0=eE/
m_{\mathrm{e}}\omega c$ was varied in the range $a_0=12$ to 72,
corresponding to a maximum laser pulse intensity of $(0.2$ to $7)\times
10^{21}$\,W/cm$^2$ and a laser power 35 to $1200$\,TW. The laser pulse was
focused on the front side of a initially cold plasma target with uniform density profile perpendicular to the $x$
direction. The electron densities were in the range from a few percents of
the electron critical density ($n_{\mathrm{c}}$) to one critical density.
For each density, the target thickness $l$ was varied from the thickness
equal to the pulse length, $L=c\tau$, up to that corresponding to almost
entire pulse depletion. The simulations were performed with the
moving-window technique in a simulation window $x\times y\times z=58\lambda
\times25\lambda\times25\lambda$ and with spatial grid steps $0.04\lambda
\times0.1\lambda\times0.1\lambda$. Typical run was about 350000 CPU hours.
\begin{figure} [!ht]
\centering{\includegraphics[width=12cm]{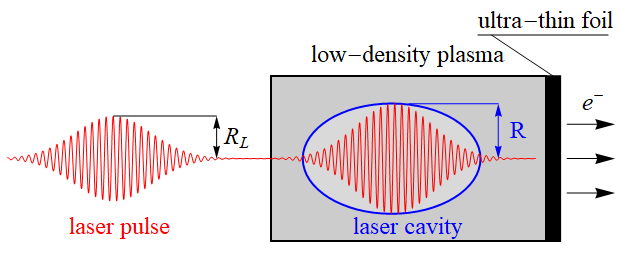}}		
	\caption{Laser--target layout}
	\label{fig1}
\end{figure}

To measure the energies of free-streaming electrons in the numerical
experiment without any effect of the transmitted laser pulse, we placed an
ultrathin overdense plasma foil of the 2\,$\mu$m-thickness and the electron density $n_{\mathrm{e}}=20 n_{\mathrm{c}}$ at the backside of the low-density target to reflect the
residual part of the laser pulse. The laser pulses considered are almost entirely depleted when hit a dense slab. Only $\sim 5$ \% of laser energy reaches it and cannot affect generated high-energy ($\gtrsim 30$ MeV) electrons at $2 \mu m$ length. The simulation laser--target layout is shown in Fig.~\ref{fig1}.

\section{Laser pulse propagation and electron generation in a subcritical plasma}\label{sec3}

\begin{figure} [!ht]
\centering{\includegraphics[width=10cm]{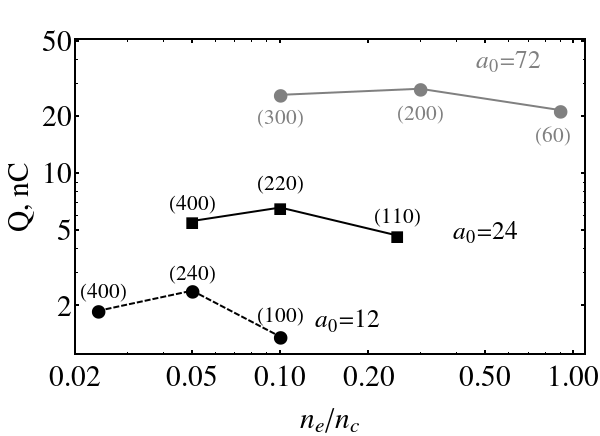}}	
	\caption{Maximum total electron charge vs.~target densities for 30\,fs
laser pulses with the amplitudes $a_0=72$ (gray dots), $a_0=24$ (black
squares) and $a_0=12$ (black dots). The number in parentheses corresponds to
the optimal target thickness in $\mu$m.}
	\label{fig2}
\end{figure}

We present the results for high-charge electron bunch generation and charge
maximization in Fig.~\ref{fig2}. For each laser pulse intensity and given
beam width, we performed several runs with different target parameters
(densities and thicknesses) to maximize the charge of accelerated electrons
with energies in excess of 30\,MeV. For each target density, we found an
optimal target thickness, $l_{opt}$, shown by the numbers (in microns) near the
resulting dots in Fig.~\ref{fig2}. This thickness, which is inversely proportional to target density, $l_{opt} \propto a_0/n_e$, is determined by the length of laser pulse depletion due to pulse etching \cite{decker} that
was used to interpret the PIC simulations \cite{Lobok,lu}.

In our simulations with a subcritical density, we found that there is an
optimal target density that allows reaching the maximum possible charge of
high-energy electrons (for definiteness, with energies higher than 30\,MeV)
with a laser pulse of a given intensity and width. This corresponds to the
condition of laser pulse propagation inside the target over several
Rayleigh lengths. The performed PIC simulations clearly confirm the
importance of matching the self-consistent laser cavity radius $R$ to the
electron plasma density and laser field amplitude \cite{gordienko,lu,ralph}.
The proposed condition for matching $R$ based on simulations for
indestructible pulse propagation in a relativistic plasma ($\gamma\sim a_0
\gg1$) is \cite{gordienko,lu,ralph}
\begin{equation}
R\simeq\alpha\frac{c}{\omega_{\mathrm{p}}}\sqrt{a_0}=
\alpha\frac{c}{\omega}\sqrt{a_0\frac{n_{\mathrm{c}}}{n_{\mathrm{e}}}}\quad{\rm or}\quad R\simeq\frac{c}{\omega}\sqrt{\frac{n_{\mathrm{c}}}{n_{\mathrm{e}}}}\left(\frac{16\alpha^4P}{P_{\mathrm{c}}}\right)^{1/6}\!,\label{eq1}
\end{equation}
where $\omega_{\mathrm{p}}$ is the electron plasma frequency
($\lambda_{\mathrm{p}}=2\pi c/\omega_{p}$), $P=E_0^2R^2c/8$ is the laser
power, $P_{\mathrm{c}}=2(m_{\mathrm{e}}c^3/r_{\mathrm{e}})(\omega^2/
\omega_{\mathrm{p}}^2)$ is the critical power for relativistic
self-focusing \cite{sun}, $P_{\mathrm{c}}\simeq17(n_{\mathrm{c}}/
n_{\mathrm{e}})$\,GW, $r_{\mathrm{e}}=e^2/m_{\mathrm{e}}c^2$ is the
classical electron radius, and $\alpha$ is a numerical factor of the
order of unity, for example, $\alpha\simeq1.12$ to 2
\cite{Lobok,gordienko,lu,ralph,Popov}.

\begin{figure} [!ht]
\centering{\includegraphics[width=16 cm]{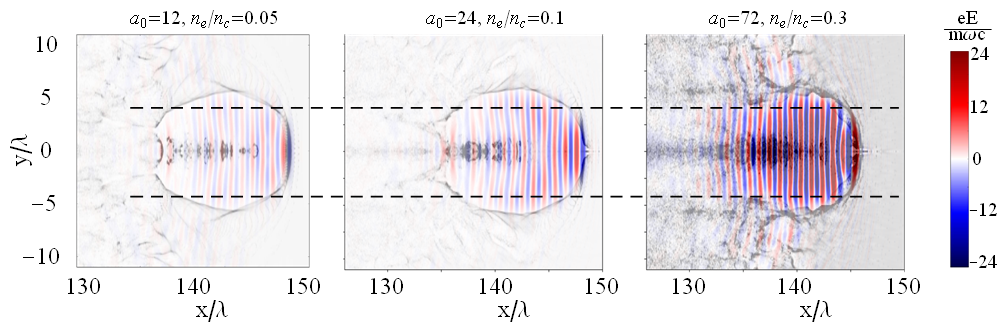}}
	\caption{Comparison of three self-trapping pulses with the initial radii
$R_L=2.4\,\mu$m and the amplitudes $a_0=12$ (left), $a_0=24$ (middle), and
$a_0=72$ (right) propagating in plasmas with the corresponding electron
densities $0.05 n_{\mathrm{c}}$, $0.1n_{\mathrm{c}}$, and
$0.3n_{\mathrm{c}}$.}
	\label{fig3}
\end{figure}

We also checked that matching condition (\ref{eq1}) corresponds to the most
stable laser pulse propagation in the self-trapping regime and to the
maximum charge of high-energy electrons by varying the laser field
amplitude. Figure \ref{fig3} clearly demonstrates the stability of the laser
pulse radius (the characteristic scale of laser intensity decrease). It
is the same, $R\simeq3.5$ to 4\,$\mu$m if the ratio $a_0/n_{\mathrm{e}}$ is
constant while the laser intensity changes by a factor of 36 and the
electron density changes by a factor of 6.

\begin{figure} [!ht]
\centering{\includegraphics[width=16 cm]{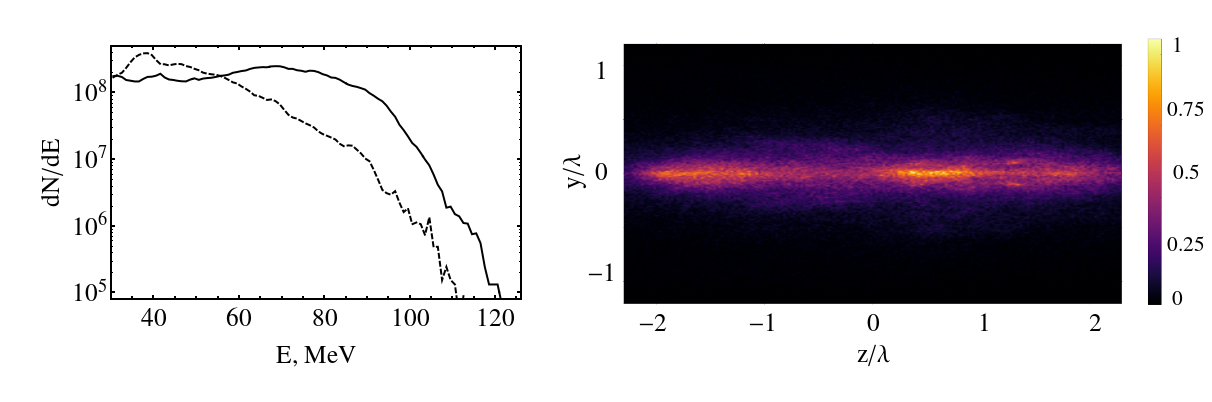}}
	\caption{Electron spectra (left panel) for the instants of 645 fs (dashed curve) and 950 fs (solid curve) and averaged over 5 check-points (from 800 fs to 950 fs) electron beam cross section (right panel) for $a_0=12$.}
	\label{fig34}
\end{figure}

For the intensity range
(0.8~to~7)${}\times10^{21}$\,W/cm$^2$, the electron energy distributions show
plateau-type spectra for the energies $\epsilon_{\min}<\epsilon_{\mathrm{e}}<
\epsilon_{\max}$, which correspondingly increase with $I$ from 20\,MeV${}<
\epsilon_{\mathrm{e}}<200$\,MeV (see. Fig. \ref{fig34}) to 50\,MeV${}<\epsilon_{\mathrm{e}}<
450$\,MeV. Such spectra are typical for stochastic electron acceleration in
the combined laser and plasma electrostatic fields \cite{Lobok,bochkarev}.
Experimental signatures that electrons can gain energy from both the
laser-excited electrostatic plasma field and the DLA mechanism have recently
been identified \cite{Shaw}. The electron beam spot size is of the order of several microns. The spot shape is elliptical. It is elongated along laser polarization direction due to the electron quiver oscillations  (see Fig. \ref{fig34}). The typical electron divergence is approximately 50 mrad, that corresponds to the emmitance of about 0.1 rad$\times \mu$m.

\section{Self-trapping regime}\label{sec4}

To understand condition (\ref{eq1}) for stable relativistic self-channeling
of a Gaussian laser beam entering a plasma, we use a commonly used model based on the
nonlinear Schr\"odinger equation (NLSE) for the complex amplitude of the
laser electric field~\cite{kovalev}. Although this model is based on a
simple stationary envelope treatment, it can shed light on the physics of
the propagation of a short laser pulse. Relativistic self-focusing is
standardly associated with the effects of the relativistic electron mass and
ponderomotive charge-displacement self-channeling, which appear in the NLSE
via the plasma nonlinear dielectric permittivity $\varepsilon_{nl}$. The
influence of both mechanisms on the propagation mode for a relativistic
laser beam has been discussed in detail in various
papers~\cite{sun,Bor_prl_92,Bor_pra_92,Bor_prl_97, fei-pre-98} and
monographs~\cite{Bor_bk_03,Shen_bk_09}. A recent work~\cite{kovalev} gives an analytical description of a self-focusing structure formation for a laser beam having a given form of the radial intensity distribution at the plasma entrance which can be plasma-vacuum interface, as in our PIC model. This work which includes the mentioned relativistic nonlinearities has taken an important step towards the solution of the boundary-value problem for the incident Gaussian light beam, that is exactly what corresponds to PIC simulations with parameter $R$ inside plasma. Strictly speaking, the theory presented in Ref.~\cite{kovalev} applies to a circularly polarized laser wave, that simplifies analytical treatment. The case of the linear polarization, which is more relevant  to  comparison with  PIC simulation, corresponds, as indicated at the end of this section, to the renormalization of the laser field amplitude. 

\begin{figure}
	\centerline{\includegraphics[width=0.98\textwidth]
		{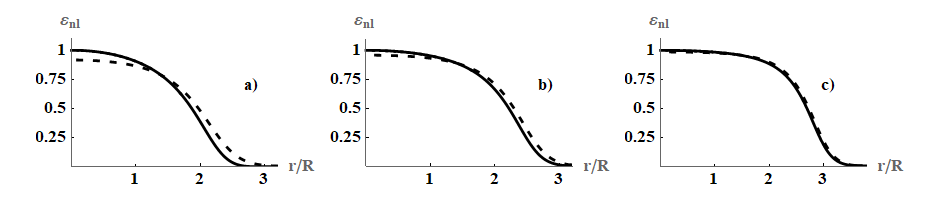}}
	\caption{Examples of the spatial nonlinear dielectric permittivities of a plasma for three sets of laser-plasma parameters: $\rho\equiv \omega_{pe} R/c=4.86$, $a_0=12$ (a), $\rho=6.88$, $a_0=24$ (b) and $\rho=11.92 $, $a_0=72$ (c) relevant to our PIC simulations. Solid curves correspond to both relativistic and charge-displacement nonlinearities and dashed curves correspond to the  relativistic nonlinearity only.}
	\label{fig4} 	
\end{figure}

One of the regimes for a weakly or modestly relativistic plasma (with
$a_0\lesssim1$) predicted and described in Ref.~\cite{kovalev} is a
self-trapping regime where a laser pulse propagates with an unchanged radius
(in reality, until depleted). In the case of interest with $a_0\gg1$, we now
derive the condition for a self-trapping regime based on a simplified
version of the analytic theory \cite{kovalev}. Such a simplification is due
to the fact that the relativistic electron nonlinearity dominates and the
effect of the charge displacement of electrons from the plasma channel, even
to the extent of complete evacuation (electron cavitation), can be neglected.
This is apparent when comparing the curves shown in Fig.~\ref{fig4} for the
dependence of the nonlinear dielectric permittivity normalized to
$\omega_{p}^{2}/\omega^2$ at the plasma boundary on the dimensionless radial
coordinate $r/R$. The very small difference between these curves argues
for using the simpler model neglecting the ponderomotive charge-displacement
effect at a relativistically high laser intensity.

\begin{figure}
\centerline{\includegraphics[width=0.6 \textwidth]{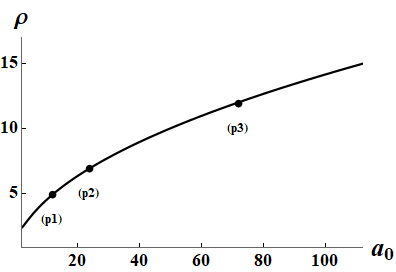}}	
	\caption{Matching radius for the self-trapping regime derived using the
analytic theory \cite{kovalev}.}
	\label{fig5} 	
\end{figure}

The self-trapping regime requires a balance of nonlinear and diffraction
effects in the relativistic plasma and corresponds to certain relation
between the plasma and laser parameters. For the simplified nonlinearity, as
discussed, this relation is $L(\rho,a_0)=0$, where the theory~\cite{kovalev}
gives
\begin{equation}
\label{eq2}
L(\rho,a_0)=(\rho^2a_0^2/2)\bigl(1+a_0^2\bigr)^{-3/2}-1.
\end{equation}
We show $\rho\equiv\omega_{pe}R/c$ as a function of $a_0$ in
Fig.~\ref{fig5}. For ultrarelativistic laser intensities, this equality
reduces to Eq.~(\ref{eq1}) with $\alpha=\sqrt{2}$. Three dots, which are
very close to the matching curve in Fig.~\ref{fig5}, indicate the parameters
used in the numerical experiment, namely, $a_0=12$, $n/n_c = 0.05$ (p1), $a_0=24$, $n/n_{\mathrm{c}}=0.1$
(p2), and $a_0=72$, $n/n_{\mathrm{c}}=0.3$ (p3), all for $R/\lambda=3.5$.
The latter corresponds to Fig.~\ref{fig3}.

The analytic solution of the NLSE for the parameters under the curve in
Fig.~\ref{fig5} shows diffraction spreading of the laser beam. Its
self-focusing occurs above this curve. But this solution should be unstable
because of filamentation, which is not an issue in the axially symmetric
stationary NLSE model. Therefore, for an ultrarelativistically intense laser,
we can expect stable pulse propagation namely for the matching curve in
Fig.~\ref{fig5}. We note that the self-trapping regime occurs not only under
the exact condition $L(\rho,a_0)=0$ but also in a small vicinity of this
curve corresponding to small positive $L(\rho,a_0)$ because for positive but
small $L(\rho,a_0)\ll1$, the solution singularity in the analytic theory
arises at a very large propagation distance $\propto1/\sqrt{L(\rho,a_0)}$.
The same holds for approaching the matching curve from the bottom, from the
side where diffraction effects dominate, i.e., for small negative values of
$L(\rho,a_0)$, which corresponds to the propagation distance
$1/\sqrt{|L(\rho,a_0)|}$.

It is interesting that the self-trapping condition $L(\rho,a_0)=0$ in
view of Eq.~(\ref{eq2}) can be derived from a trivial perturbation theory
approximation for laser ray evolution equations in the near-axis region
near the plasma boundary. Moreover, this condition also arises in
heuristic theoretical models based on \textit{ad hoc} assumptions on the
laser beam structure inside the plasma (see, e.g., \cite{sun}). But the
procedure used in \cite{kovalev} gives a self-consistent justified result
for $L(\rho,a_0)$ because it gives the complete spatial distribution of the
laser beam electric field, which is not limited by the proximity to the
plasma entrance area nor the vicinity of the beam axis and does not use any
initial assumptions concerning the beam structure. Our theory also removes
the contradiction between other heuristic models, aberrationless and moment
approximations \cite{walia}.

We also comment regarding the effect of the polarization of laser light
\cite[p.~37]{Bor_bk_03} (see also \cite{sen-opt-13}). For a linearly polarized laser wave, the value of
$a_0$ in the relativistic gamma-factor expression should be replaced with
$a_0/\sqrt{2} $. Hence, in relation~(\ref{eq2}) corresponding to the theory
for circularly polarized light \cite{kovalev}, we should make such a
replacement, which results in a $2^{1/4}$ increase of $\alpha$, i.e.,
$\alpha= 2^{3/4}$ for a linearly polarized laser wave.

Finally, we comment on the applicability limit of the stationary (in
time) NLSE approach based on the approximation of a slow evolution of the
envelope amplitude of the laser electric field, which we believe can be
used to qualitatively derive the matching condition obtained in PIC
simulations. For rather long laser pulses, a slow time evolution can be
described by assuming a parametric dependence of the beam amplitude on the
time variable $t-z/v_{g}$, as was done in~\cite[Chap.~9.1.1, p.~136]{Bor_bk_03}.
In this approach, the laser pulse is treated as consisting of transverse
``slices,'' and the propagation of each slice can be simulated independently.
The discussed approach cannot take into account fast laser beam evolution
at the plasma entrance. That is why we have used the modified beam radius, $R$, from PIC simulations as an input parameter for the theoretical model.
Of course, for a precise model of the ultrashort pulse physics, the
nonlinear wave model with second-order time and spatial derivatives must be
used instead of the NLSE approach \cite[Chap.~9.2, p.~139--141]{Bor_bk_03}.

\section{Discussion and conclusions}\label{sec5}

\begin{figure} [!ht]
\centering{\includegraphics[width=12cm]{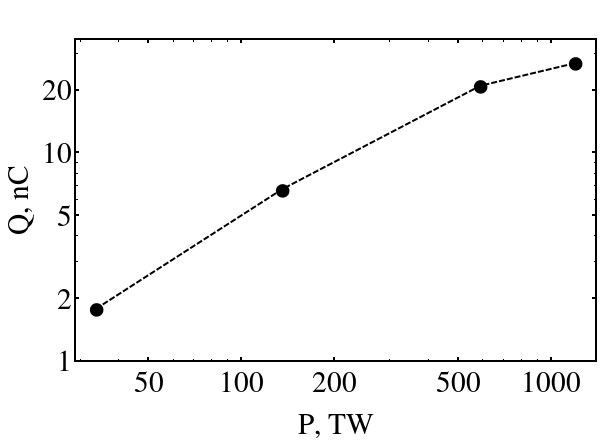}}
	\caption{Dependence of the maximum total electron charge on the laser power
for a 30\,fs pulse.}
	\label{fig6}
\end{figure}

We have here focused on studying how to accelerate as many electrons as
possible, which is incompatible with the usual goal of obtaining
quasimonoenergetic electrons. Nevertheless, our case is somewhat close to
the latter because we observe plateau-type electron spectra. Our PIC
simulations showed that joint DLA and electrostatic acceleration of
electrons in a laser cavity propagating in the self-trapping regime produces
multi-nC hundred-MeV electron bunches at ultrarelativistic intensities of
tightly focused laser pulses and subcritical plasma densities. The
dependence of the maximum total electron charge on the laser power for a
30\,fs pulse in our PIC simulations is shown in Fig.~\ref{fig6}.

We also demonstrated that the analytic theory of relativistic self-focusing
of an initially Gaussian laser beam \cite{kovalev} in the considered case of
ultrarelativistic laser intensities $a_0\gg1$ predicts that a self-trapping
regime of laser pulse propagation exactly corresponds to condition~(\ref{eq1}).
In this regime, the cavity adjusts the initial laser beam spot to a
self-consistent size that evolves very little during propagation. The laser
cavity propagates over many Rayleigh lengths and effectively accelerates a
large number of electrons, i.e., condition~(\ref{eq1}) for subcritical
plasmas is essential for producing electron beams with a high total charge.
For radii larger than that given by Eq.~(\ref{eq1}), the laser pulse
propagation is filamentary and unstable in the transverse direction. For
smaller radii, a laser pulse spreads because the diffraction is too high for
self-channeling.

Since proposals for using intense ultrashort laser pulses to trigger nuclear
reactions first appeared \cite{bychenkov,ledingham}, this issue has been
overgrown with numerous original nuclear applications. Electron acceleration
by a laser-excited plasma wakefield has been used to produce gamma-rays by
passing through high-Z material converters. This technique has distinct
advantages over direct laser irradiation of solid targets because the
electron source is small. One of the first experiments on generating
Bremsstrahlung gamma rays from a LWFA was reported in Ref.~\cite{edwards}.

For high photon energies beyond several MeV, generation by Bremsstrahlung
is most attractive. This is to be expected from the electron source proposed
here. Correspondingly, a gamma source should be small and bright because the
total electron charge is high. This makes the considered electron source
potentially useful for gamma-ray radiography with high spatial resolution.
We also believe that a high-charge electron source can be beneficial for
tabletop photonuclear reactions inducing photofission, generating neutrons,
generating electron--positron pairs, and even possibly producing light
mesons. The last has already been demonstrated experimentally
\cite{schumaker}. We consider these matters important for our future study.

\section{Acknowledgments}
	
This work was supported by the Russian Science Foundation (Grant
No.~17-12-01283).

\end{document}